\documentclass[cits,a4paper]{PoS}

\usepackage{amsmath}

\newcommand{\bra}[1]{\ensuremath{\langle #1 |}}   
\newcommand{\ket}[1]{\ensuremath{| #1 \rangle}}   
\renewcommand{\vec}[1]{{\bf #1}}

\title{Leptophilic Dark Matter in Direct Detection Experiments and in the Sun }
\ShortTitle{Leptophilic Dark Matter in Direct Detection Experiments and in the Sun }

\author{\speaker{Joachim KOPP} \\
        Fermilab, PO Box 500, Batavia, IL 60510, USA \\
        E-mail: \email{jkopp@fnal.gov}}

\author{Viviana NIRO\\
        Max-Planck-Institute for Nuclear Physics,
        PO Box 103980, 69029 Heidelberg, Germany \\
        E-mail: \email{niro@mpi-hd.mpg.de}}

\author{Thomas SCHWETZ\\
        Max-Planck-Institute for Nuclear Physics,
        PO Box 103980, 69029 Heidelberg, Germany \\
        E-mail: \email{schwetz@mpi-hd.mpg.de}}

\author{Jure ZUPAN\\
        Faculty of Mathematics and Physics, Univ.\ of Ljubljana,
        Jadranska 19, 1000 Ljubljana, Slovenia \\
        Department of Physics, University of Cincinnati, Cincinnati, Ohio  
        45221, USA \\
        Josef Stefan Institute, Jamova 39, 1000 Ljubljana, Slovenia \\
        SISSA, Via Bonomea 265, I 34136 Trieste, Italy\\
        E-mail: \email{jure.zupan@cern.ch}}

\abstract{Dark matter interacting predominantly with leptons instead of nuclear
matter has received a lot of interest recently. In this talk, we investigate
the signals expected from such 'leptophilic Dark Matter' in direct detection
experiments and in experiments looking for Dark Matter annihilation into
neutrinos in the Sun. In a model-independent framework, we calculate the
expected interaction rates for different scattering processes, including
elastic and inelastic scattering off atomic electron shells, as well as
loop-induced scattering off atomic nuclei. In those cases where the last effect
dominates, leptophilic Dark Matter cannot be distinguished from conventional
WIMPs. On the other hand, if inelastic scattering off the electron shell
dominates, the expected event spectrum in direct detection experiments is
different and would provide a distinct signal. However, we find that the
signals in DAMA and/or CoGeNT cannot be explained by invoking leptophilic DM
because the predicted and observed energy spectra do not match, and because of
neutrino bounds from the Sun.}

\FullConference{Identification of Dark Matter 2010\\
                July 26 - 30 2010\\
                University of Montpellier 2, Montpellier, France}

\allowdisplaybreaks

\begin{document}

\section{Introduction}
\label{sec:intro}

The possibility that Dark Matter interacts predominantly with leptons has
recently received a lot of attention~\cite{Fox:2008kb, Ibarra:2009bm,
Spolyar:2009kx, Chun:2009zx, Kyae:2009jt, Bi:2009uj, Davoudiasl:2009dg,
Cohen:2009fz, Farzan:2010mr, Haba:2010ag, Erkoca:2010vk}, in particular in the
context of cosmic ray anomalies~\cite{Adriani:2008zr, Abdo:2009zk,
Hooper:2010mq} which could be due to Dark Matter annihilation. The
phenomenology of leptophilic Dark Matter in direct detection
experiments~\cite{Bernabei:2007gr, Fox:2008kb, Dedes:2009bk, Kopp:2009et} is
somewhat less well explored, even though it has been
noted~\cite{Bernabei:2007gr} that a scattering process in which the recoil
energy is transferred to electrons could explain the annual modulation signal
observed in DAMA~\cite{Bernabei:2008yi, Bernabei:2010mq} while remaining
consistent with constraints from other experiments which treat electron recoils
as background. For similar reasons, one might also hope to explain the CoGeNT
signal~\cite{Aalseth:2010vx} by invoking leptophilic Dark Matter.

In this talk, which is mainly based on Ref.~\cite{Kopp:2009et}, we study in
detail the expected direct detection signals from leptophilic Dark Matter
scattering. We will introduce our formalism, based on an effective field theory
description of Dark Matter scattering, in sec.~\ref{sec:model}, and then
proceed to a discussion of the four different classes of processes that can
occur when a leptophilic weakly interacting massive particle (WIMP) interacts
in a detector: WIMP-electron scattering, elastic and inelastic WIMP-atom
scattering, and loop-induced WIMP-nucleus scattering. In sec.~\ref{sec:sim}, we
present exclusion limits on leptophilic Dark Matter from various direct
detection experiment, and in sec.~\ref{sec:sun} we supplement these results
with limits on leptophilic WIMP annihilation into neutrinos in the Sun. We
summarize our results and conclude in sec.~\ref{sec:conclusions}.

\section{Leptophilic Dark Matter}
\label{sec:model}

Interactions between a Dark Matter (DM) fermion $\chi$ and charged leptons
$\ell$ can be introduced in a model-independent way by considering the
effective operator
\begin{align}
  \mathcal{L}_{\rm eff} = \sum_i G \, (\bar\chi \Gamma_\chi^i \chi) \,
                                      (\bar\ell \Gamma_\ell^i \ell)
  \qquad\text{with}\qquad G = \frac{1}{\Lambda^2}\,,
\end{align}
where $\Lambda$ is the UV-completion scale of the effective field theory,
and $\Gamma_\chi$, $\Gamma_\ell$ are Lorentz tensors. In principle, one
can consider the following Lorentz structures:
\begin{align}
  \renewcommand{\arraystretch}{1.5}
  \begin{array}{l@{\qquad}l@{\quad}l}
    \text{scalar (S) / pseudoscalar (P):} &
    \Gamma_\chi = c^\chi_S + ic^\chi_P\gamma_5, &
    \Gamma_\ell = c^\ell_S + ic^\ell_P\gamma_5 \,, \\
    \text{vector (V) / axial vector (A):} &
    \Gamma_\chi^\mu = (c^\chi_V + c^\chi_A\gamma_5)\gamma^\mu, &
    \Gamma_{\ell\mu} = (c^\ell_V + c^\ell_A\gamma_5)\gamma_\mu \,,\\
    \text{tensor (T) / axial tensor (AT):} &
    \Gamma_\chi^{\mu\nu} = (c_T + ic_{AT} \gamma_5)\sigma^{\mu\nu}, &
    \Gamma_{\ell\mu\nu} = \sigma_{\mu\nu} \,.
  \end{array}
\end{align}
However, it is straightforward to show~\cite{Kopp:2009et} that for many
of these operators, the low-energy WIMP-electron scattering cross section
is proportional to $v^2$, where $v$ is the WIMP velocity. Since, in units
of the speed of light, $v^2 \sim \mathcal{O}(10^{-6})$, these terms are
negligible in direct detection experiments unless all unsuppressed terms
are absent and the cutoff scale $\Lambda$ is very low. While this possibility
cannot be excluded in a model-independent way, it is ruled out in many
concrete DM models, which is why in most studies only the unsuppressed
operators
\begin{align}
  \renewcommand{\arraystretch}{1.5}
  \begin{array}{l@{\quad}l@{\qquad\quad}l@{\quad}l}
    S \otimes S: & G \, (\bar\chi \chi) \, (\bar\ell \ell) \,, &
    V \otimes V: & G \, (\bar\chi \gamma^\mu \chi) \, (\bar\ell \gamma_\mu \ell) \,, \\
    A \otimes A: & G \, (\bar\chi \gamma^\mu \gamma^5 \chi) \,
                        (\bar\ell \gamma_\mu \gamma^5 \ell) \,, &
    T \otimes T: & G \, (\bar\chi \sigma^{\mu\nu} \chi) \, (\bar\ell \sigma_{\mu\nu} \ell) \,
  \end{array}
\end{align}
are considered. Here, we will in particular focus on $V \otimes V$ and $A
\otimes A$ operators since we will see that, as far as the direct detection
phenomenology is concerned, operators with other Lorentz structures are
qualitatively similar to either of the two.

\begin{figure}
  \begin{center}
    \includegraphics[width=\textwidth]{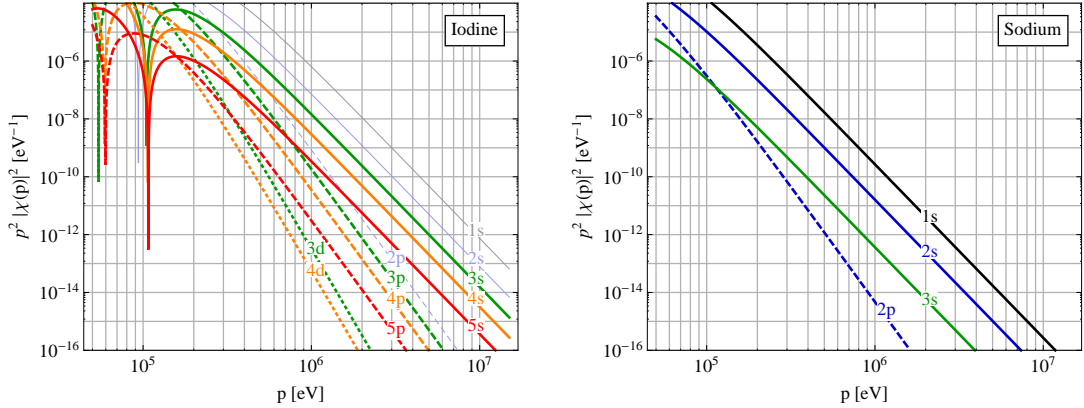}
  \end{center}
  \vspace{-0.7cm}
  \caption{The momentum space electron wave functions of iodine and sodium. Thick
    colored curves correspond to shells that contribute to WIMP-electron
    scattering in DAMA, while thin light curves correspond to electrons
    that are too tightly bound to be separated from the atom in a WIMP
    interaction at DAMA energies. The approximate wave functions shown here
    are taken from ref.~\cite{Bunge:1993a}. They do not include
    relativistic corrections (which can lead to flattening at high momentum)
    or multi-electron correlations.}
  \label{fig:wf}
\end{figure}

A leptophilic WIMP can interact in a detector in four different ways:

\emph{(i) WIMP-electron scattering.}
If the WIMP interacts with a weakly bound electron (i.e.\ the energy
transferred to the electron is much larger than its binding energy to the
atomic nucleus), the electron will be kicked out of the atom to which it is
bound, while the atom remains at rest.  The typical electron recoil energy in
processes of this type if of order $m_e v^2 \lesssim 1$~eV, far below the
$\mathcal{O}(\text{keV})$ detection thresholds of DM direct detection
experiments. However, since the electron is initially in a bound state, there
is a small probability that it enters the interaction with a very high initial
state momentum.  In this case the kinematics is different, and
$\mathcal{O}(\text{keV})$ recoil energies are possible, though unlikely. More
precisely, the differential event rate for axial-vector WIMP-electron
scattering is given by
\begin{align}
  \frac{dR}{dE_d} \simeq
  \frac{3 \rho_0 m_e G^2}{4\pi m_{\rm A} m_\chi} \,
    \sum_{nl} \sqrt{2 m_e (E_d - E_{B,nl})} \, (2l+1)
    \int\!\frac{dp\,p}{(2\pi)^3} \, |\chi_{nl}(p)|^2 \,
    I(v_{\rm min}) \,,
  \label{eq:dRdE-WES}
\end{align}
where $m_\chi$ is the WIMP mass, $\rho_0 \sim 0.3$~GeV cm$^{-3}$ is the
local DM density $m_{\rm A}$ is the mass of the target atom, $E_d$
is the observed electron recoil energy, $E_{B,nl}$ is the
binding energy of the $(n,l)$ atomic shell, and $\chi_{nl}(p)$ is the
radial part of the momentum-space wave function of that shell, normalized
according to $\int\!dp \, (2\pi)^{-3} p^2 |\chi_{nl}(p)|^2 = 1$.
The function $I(v_{\rm min})$ is defined by $I(v_{\rm min}) \equiv \int\!d^3v
\;\; v^{-1} f(\vec{v}) \, \theta(v - v_\mathrm{min})$,
with the WIMP velocity distribution $f(\vec{v})$ and the minimum
WIMP velocity required to obtain a recoil energy $E_d$ given by
$v_{\rm min} \approx E_d / p + p / 2 m_\chi$. To arrive at
eq.~\eqref{eq:dRdE-WES}, we have evaluated the Feynman diagram of
WIMP-electron scattering, taking into account the modified kinematics
for bound systems, and replacing the usual plane wave initial states
with the appropriate bound state wave functions.
Comparing eq.~\eqref{eq:dRdE-WES} to the rate for conventional
WIMP-nucleus scattering through axial-vector couplings in a nucleophilic
model,
\begin{align}
  \frac{dR^0}{dE_d} \simeq
    \frac{3 \rho_0 G^2}{2 \pi m_\chi} \, I(v_{\rm min}^0) \qquad\text{with}\qquad
  v_{\rm min}^0 = \frac{m_\chi + m_N}{m_\chi} \sqrt{\frac{E_d}{2 m_N}} \,,
  \label{eq:dRdE-conventional}
\end{align}
we find that \eqref{eq:dRdE-WES} is suppressed compared to
\eqref{eq:dRdE-conventional} by a factor of order $m_e/m_A \times \sqrt{2 m_e
E_d} \, p^2 |\chi_{nl}(p)|^2$.  For $m_\chi \gtrsim 10$~GeV, the relevant
values for $p$ (i.e.\ those which lead to the smallest possible $v_{\rm min}$
and hence to the largest possible value of $I(v_{\rm min})$) are $p \sim
\sqrt{2 m_\chi E_d} \gtrsim 10$~MeV/c.  In fig.~\ref{fig:wf}, we plot $p^2
|\chi_{nl}(p)|^2$ (which is proportional to the momentum distribution of the
electron) as a function of $p$. We observe that $p^2 |\chi_{nl}(p)|^2$ is
extremely small in the relevant momentum region, so that $dR/dE_d$ is hugely
suppressed compared to $dR^0/dE_d$ for similar values of $G^2$.

\emph{(ii) Elastic WIMP-atom scattering.}
If a WIMP interacts with one of the strongly bound inner electrons of one of
the target atoms, the energy transfer is not sufficient to overcome the
electron binding energy, and the recoil will be taken up by the atom as a
whole. Experimentally, such events would resemble conventional nuclear recoils.
However, it turns out that elastic WIMP-atom scattering is always subdominant
compared to other processes~\cite{Kopp:2009et}, so we will not consider it
further here.

\emph{(iii) Inelastic WIMP-atom scattering.}
For scattering processes in which the energy transfer is comparable to
the binding energy of the target electron, the electron may be excited
to a less strongly bound state, but still remain bound to the nucleus.
In that case, as for elastic WIMP-atom scattering, the recoil momentum
will be taken up by the atom as a whole. The event rate for inelastic
WIMP-atom scattering is proportional to
\begin{align}
  \sum_{n'l'm'} \sum_{nlm} |\bra{n'l'm'} e^{i (\vec{k} - \vec{k'}) \vec{x}} \ket{nlm} |^2 \,,
\end{align}
where $(n, l, m)$ and $(n', l', m')$ are the initial and final state quantum
numbers of the electron and $\vec{k}$, $\vec{k}'$ are the initial and final
WIMP momenta.  Since the electron wave functions are tiny at the large momenta
required in direct detection experiments, these matrix elements are tiny.
Numerically, it turns out that inelastic WIMP-atom scattering is subdominant
compared to WIMP-electron scattering~\cite{Kopp:2009et}. However, in
experiments that reject pure electron recoils as background, it may be the
dominant contribution to the signal since it resembles WIMP-nucleus scattering.

\begin{figure}
  \vspace{-0.3cm}
  \begin{center}
    \includegraphics[height=2.4cm]{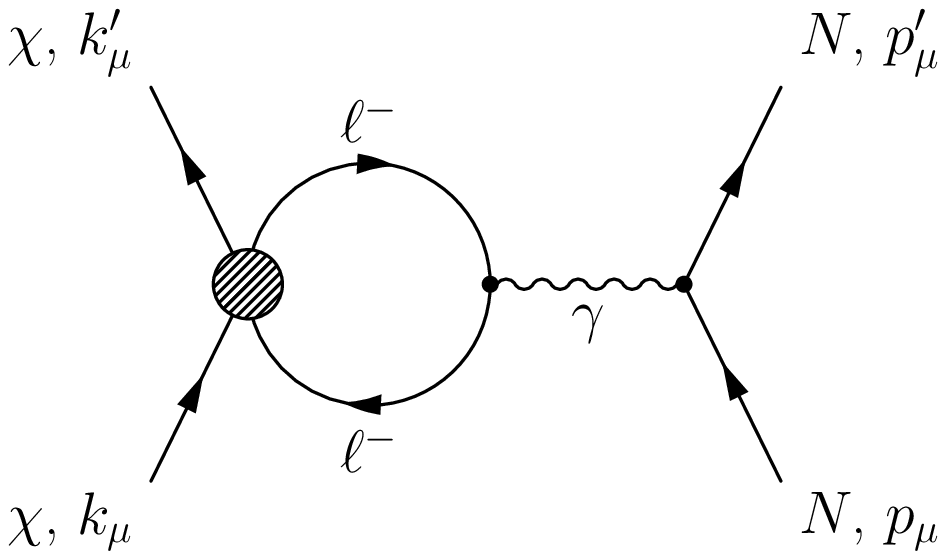} \quad
    \includegraphics[height=2.4cm]{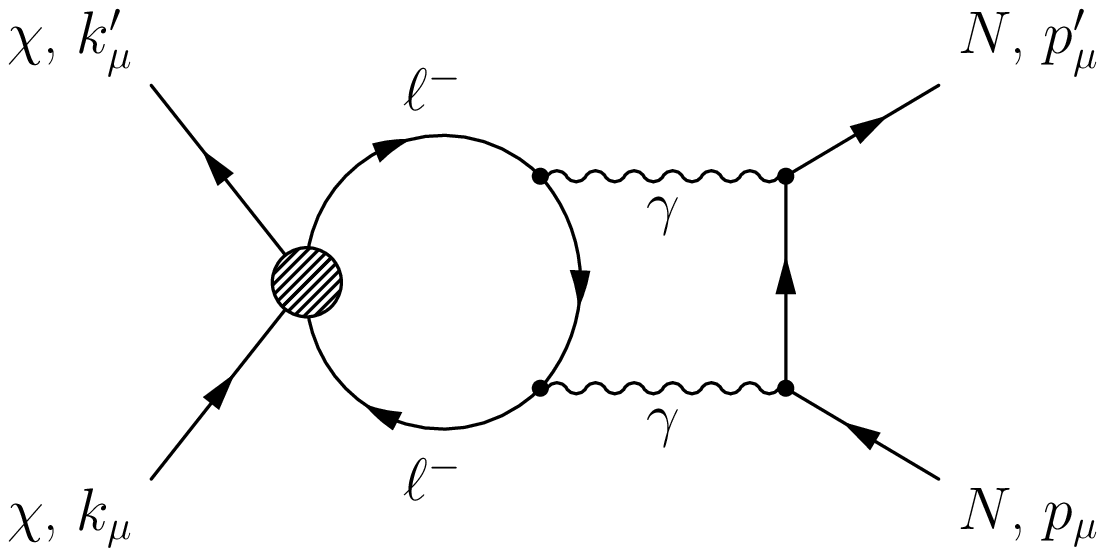}\quad
    \includegraphics[height=2.4cm]{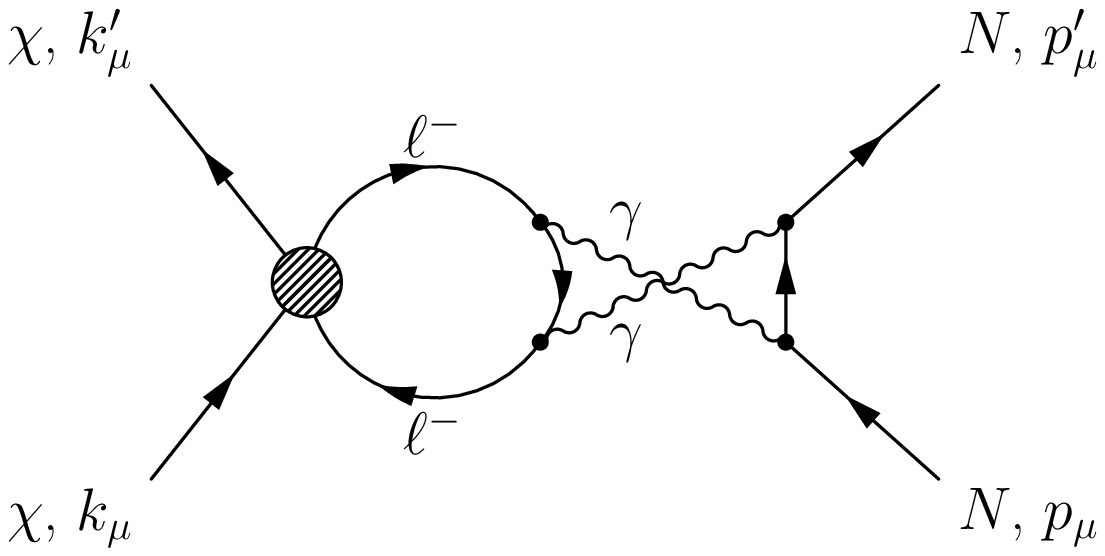}
  \end{center}
  \vspace{-0.3cm}
  \caption{DM-nucleus interaction induced by a charged lepton loop and photon
    exchange.}
  \label{fig:loop}
\end{figure}

\emph{(iv) Loop-induced WIMP-nucleus scattering.}
Even though tree level WIMP-nucleus scattering is forbidden in
our leptophilic scenario, it may be induced at the loop level through the
diagrams shown in fig.~\ref{fig:loop}. While these diagrams are suppressed by
one (1-loop) or two (2-loop) powers of $\alpha Z$, multiplied by a loop factor,
compared to tree-level WIMP-nucleus scattering in conventional nucleophilic DM
models, the suppression is much less severe than that of WIMP-electron
scattering \emph{(i)} and WIMP-atom scattering \emph{(ii)}, \emph{(iii)}.
This leads to the conclusion that, \emph{whenever WIMP-nucleus
scattering is possible, be it only at the loop-level, it will be the dominant
process in direct detection experiments.} This in particular means that
leptophilic DM can only reconcile the DAMA and CoGeNT signals with the
null observations from other experiments if the loop-diagrams from
fig.~\ref{fig:loop} vanish. This is, for example, the case for $A \otimes A$
couplings between DM and electrons.

We summarize this section with a rough numerical estimate for the relative rates
of WIMP-atom scattering \emph{(i), (ii)} (WAS), WIMP-electron scattering \emph{(iii)}
(WES), and loop-induced WIMP-nucleus scattering \emph{(iv)} (WNS)~\cite{Kopp:2009et}:
\begin{align}
  R^{\rm WAS} : R^{\rm WES} : R^{\rm WNS} \sim \sim 10^{-17} : 10^{-10} : 1 \,.
\end{align}

\section{Exclusion limits on leptophilic Dark Matter from direct detection experiments}
\label{sec:sim}

\begin{figure}
  \vspace{-0.5cm}
  \begin{tabular}{cc}
    \includegraphics[height=5.5cm]{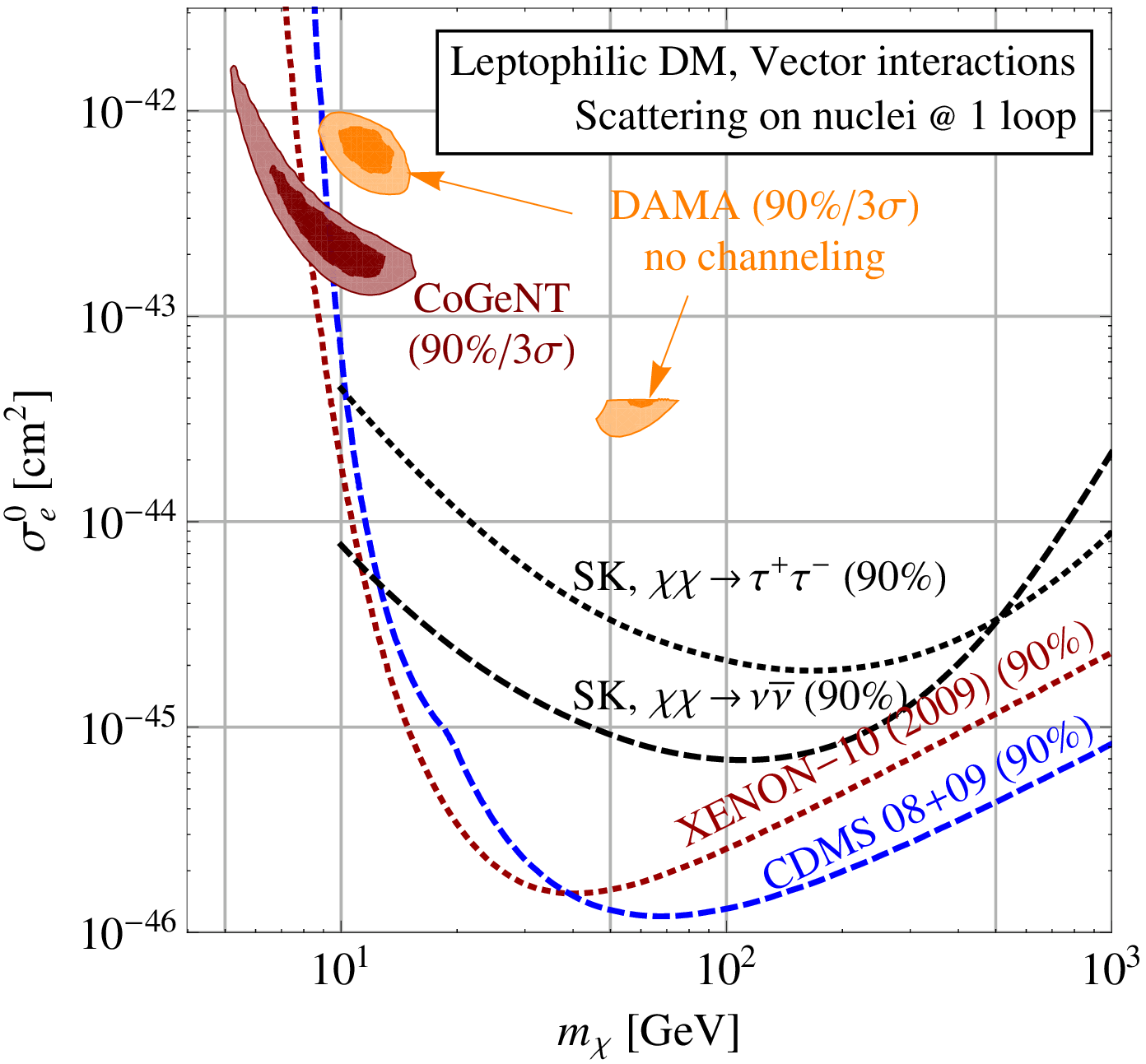} &
    \includegraphics[height=5.5cm]{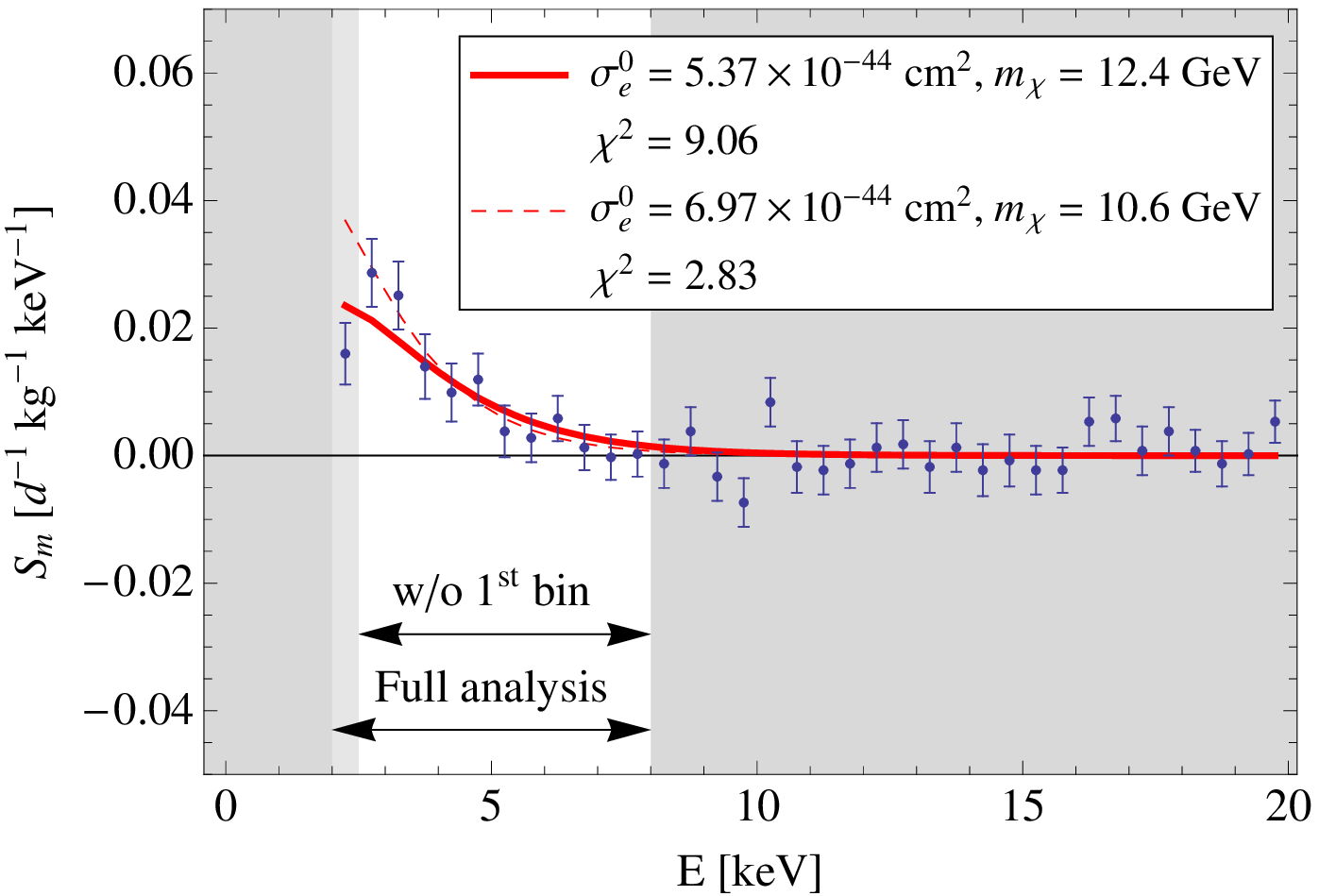} \\[-0.1cm]
    \includegraphics[height=5.5cm]{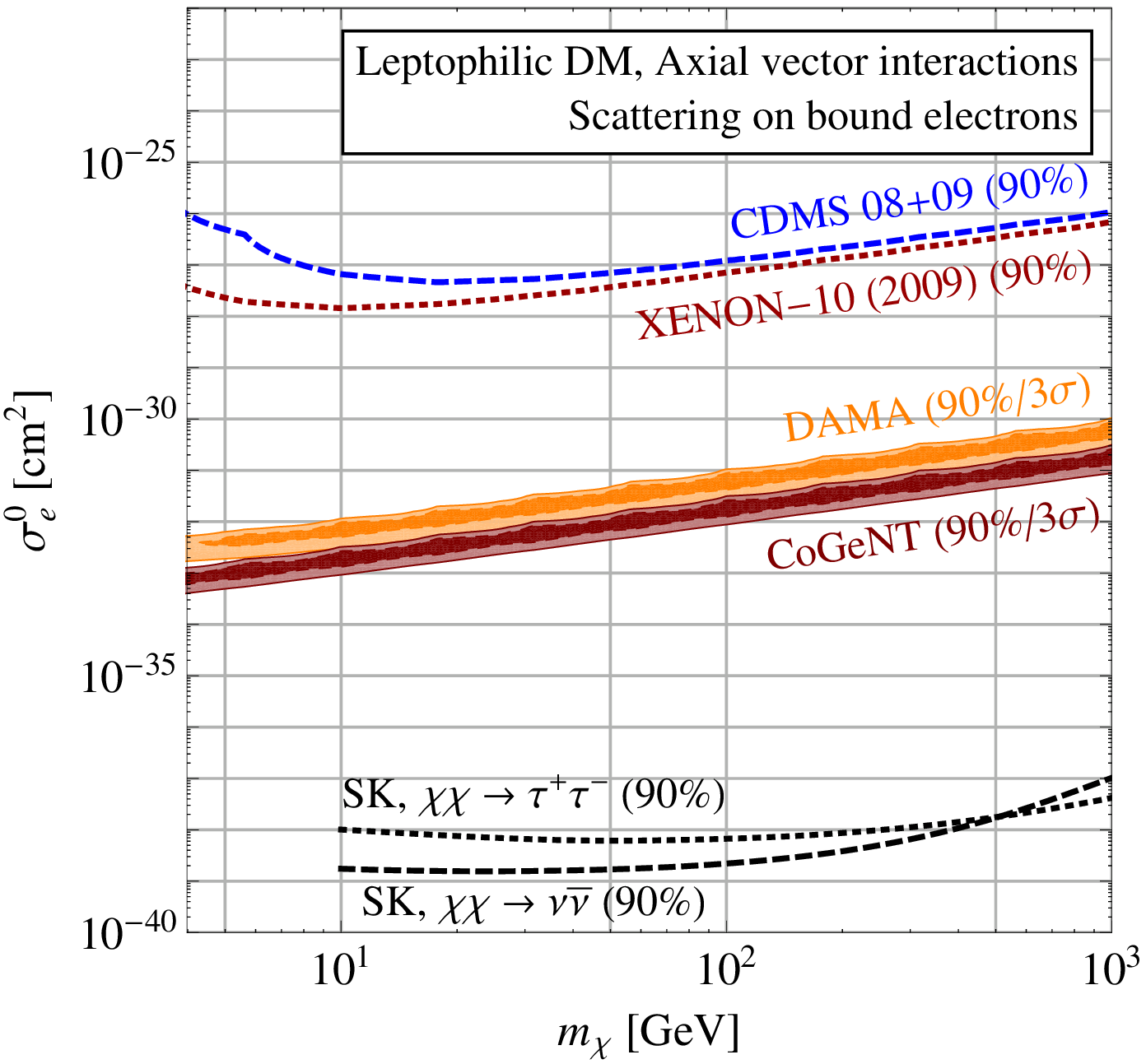} &
    \includegraphics[height=5.5cm]{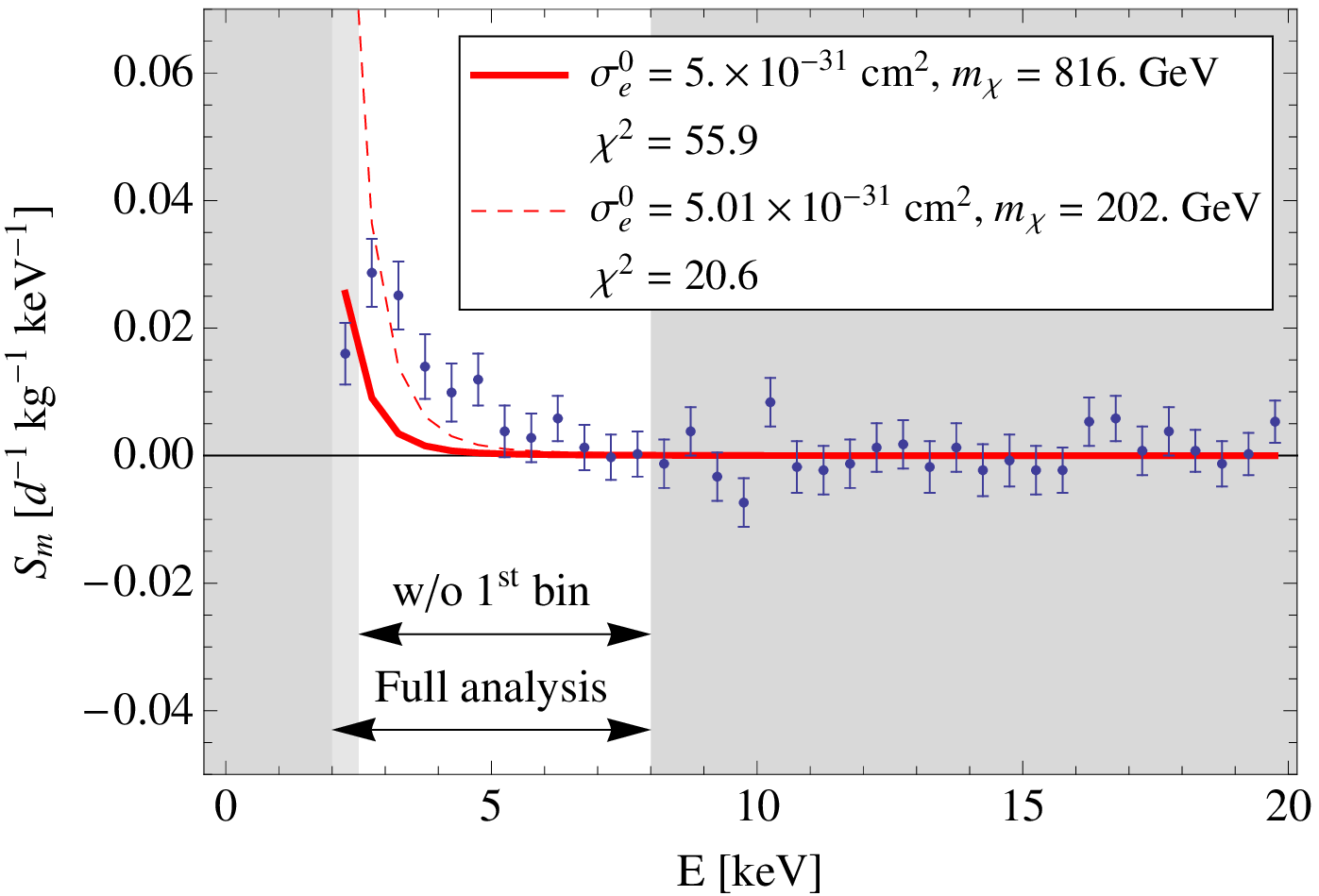}
  \end{tabular}
  \vspace{-0.1cm}
  \caption{Left: CDMS, XENON-10, and Super-Kamiokande exclusion limits
    and DAMA/CoGeNT-favored values for the WIMP mass $m_\chi$ and the WIMP--fre
    electron scattering cross section $\sigma_e^0$. Right: comparison of the
    observed annual modulation spectrum in DAMA to the prediction for leptophilic
    DM. The solid curve has been fitted to the DAMA data from 2--8~keV, while for the
    dashed curve, the first energy bin has been neglected. The top row of
    panels is for $V \otimes V$ interactions, while the bottom row is for
    $A \otimes A$ interactions.}
  \label{fig:exclusion}
\end{figure}

Let us now consider experimental constraints on leptophilic DM from direct
detection experiments. In fig.~\ref{fig:exclusion}, we show constraints on the
WIMP mass and the WIMP--free electron cross section $\sigma_e^0$ derived from
CDMS~\cite{Ahmed:2008eu}, XENON-10~\cite{Angle:2007uj},
CoGeNT~\cite{Aalseth:2010vx}, and DAMA~\cite{Bernabei:2008yi} data. We also
compare the DAMA annual modulation spectrum to the signal predicted for
leptophilic DM. We see that for $V \otimes V$ interactions, the spectral fit to
the DAMA data is good, but the tension between DAMA/CoGeNT and CDMS/XENON-10 is the
same as in non-leptophilic models. This is easily understandable because in the
$V \otimes V$ case, the loop diagrams fig.~\ref{fig:loop} are non-zero, so the
dominant signal in all experiments is due to WIMP-nucleus scattering.  On the
other hand, for $A \otimes A$ interactions, WIMP-nucleus scattering is absent,
so only the much weaker inelastic WIMP-atom scattering contributes to the
CDMS/XENON-10 exclusion limits, while the DAMA/CoGeNT signals are explained by the
less-suppressed WIMP-electron scattering. Note that the analysis of electron
background events in CDMS, ref.~\cite{Ahmed:2009rh}, could be used to improve
the CDMS limit, even though it would still be consistent with the DAMA/CoGeNT-favored
parameter region. A careful analysis of the extremely low electron background
in XENON-100~\cite{Aprile:2010um}, however, may rule out that region.  In any
case, the fit to the DAMA modulation spectrum (and the fit to the spectrum
of excess events in CoGeNT) is very poor in the case of $A
\otimes A$ couplings because the steep decrease of the electron wave functions
at high momentum (fig.~\ref{fig:wf}) leads to a too steeply decreasing
modulation spectrum. This rules out leptophilic DM with $A \otimes A$ couplings
as an explanation for the DAMA and/or CoGeNT signals. Since, as far as direct
detection experiments are concerned, $V \otimes V$ and $A \otimes A$
interactions encompass all phenomenologically different types of leptophilic DM
models, we conclude that leptophilic DM cannot reconcile DAMA and CoGeNT with
other experiments.

\section{Neutrinos from Dark Matter annihilation in the Sun}
\label{sec:sun}

In addition to constraints from direct detection experiments, we have also
considered DM capture and annihilation in the Sun, which may lead to detectable
neutrino signals. Since scattering on the free electrons in the Sun is
sufficient for a WIMP to be captured, DM capture in the Sun does not receive
the same suppression as WIMP-electron scattering observable in direct detection
experiments. Still, in those cases where WIMP-nucleus scattering is allowed, it
is the dominant capture reaction. For direct detection, it was sufficient to
assume DM couplings to electrons (which would not lead to annihilation of the
captured WIMPs into neutrinos), but it is very natural to assume that
interactions with electrons are accompanied by interactions with other leptons.
Even without that additional assumption, annihilation into neutrinos can be
induced by loop diagrams similar to those shown in fig.~\ref{fig:loop}, by a
diagram in which two DM particles annihilate into virtual electrons which then
exchange a $W$ boson and turn into neutrinos, or by $W/Z$
radiation~\cite{Bell:2010ei}.  On the other hand, neutrino signals from DM
annihilation can be absent if there exists a particle-antiparticle asymmetry in
the DM sector. If we neglect this possibility, we can derive constraints on
leptophilic DM from Super-Kamiokande data~\cite{Desai:2004pq} (black curves in
fig.~\ref{fig:exclusion}). These constraints are comparable to direct detection
constraints when WIMP-nucleus scattering dominates, but much stronger than
direct detection constraints when WIMP-electron scattering is most important.
Thus, even though Super-Kamiokande limits are not as model-independent as
direct detection constraints, they strongly support our conclusion that
leptophilic DM cannot explain the DAMA signal.

\section{Conclusions}
\label{sec:conclusions}

In conclusion, we have studied the phenomenology of the well-motivated
leptophilic Dark Matter scenario in direct detection experiments and have
made detailed predictions for the observable signals. In particular, we
have classified leptophilic DM interactions into elastic WIMP-atom scattering,
inelastic WIMP-atom scattering, WIMP-electron scattering, and loop-induced
WIMP-nucleus scattering, with the first one having the smallest cross section
and the last one the largest, unless the relevant loop-diagrams are forbidden
by symmetry arguments. We have then computed model-independent constraints
on the parameter space of leptophilic DM from CDMS, XENON-10, CoGeNT,
and DAMA data, and slightly model-dependent constraints from Super-Kamiokande
data on DM-induced neutrino signals from the Sun. While our
study shows that leptophilic DM has a rich and interesting phenomenology,
we have also seen that it cannot explain the DAMA/CoGeNT signals while
remaining consistent with null results from other experiments.

\acknowledgments
It is a pleasure to thank Marco Cirelli, Nicolao Fornengo, Roni Harnik,
Alexander Merle, Thomas E.\ J.\ Underwood, and Andreas Weiler for very useful
discussions, and the organizers of the IDM~2010 conference for a very enjoyable
and productive meeting.  This work was in part supported by the
Sonderforschungsbereich TR~27 of the Deutsche Forschungsgemeinschaft.  JK
received support from the Studienstiftung des Deutschen Volkes.  Fermilab is
operated by Fermi Research Alliance, LLC under Contract No.~DE-AC02-07CH11359
with the US DOE.

\providecommand{\bysame}{\leavevmode\hbox to3em{\hrulefill}\thinspace}

\end{document}